\documentclass{ws-procs9x6}

\begin{document}

\title{A Perturbative QCD Based Study of Polarized Nucleon Structure
in the Transition
Region and Beyond: ``Quarks, Color Neutral Clusters, and Hadrons''}

\author{S. LIUTI }

\address{Department of Physics \\
382, McCormick Rd., \\ 
Charlottesville, VA 22904-4714, USA\\ 
E-mail: sl4y@virginia.edu}

\author{N. BIANCHI and A. FANTONI}

\address{INFN, Laboratori Nazionali di Frascati \\ 
Via Enrico Fermi 40, \\
00044 Frascati (Roma) Italy\\
E-mail: fantoni@lnf.infn.it, bianchi@lnf.infn.it}  

\maketitle

\abstracts{A large fraction of the world data on both polarized 
and unpolarized inclusive $ep$ 
scattering at large Bjorken $x$ lies in the resonance region where 
a correspondence with the deep inelastic regime,
known as Bloom and Gilman's duality, 
was observed.  
Recent analyses of the $Q^2$ dependence of the data 
show that parton-hadron duality 
is inconsistent with the twist expansion 
at low values of the final state 
invariant mass.
We investigate the nature of this disagreement, and we interpret its
occurrence in terms of 
contributions from non partonic degrees of freedom in a preconfinement
model.}

\section{Introduction}
Parton-hadron duality, or the idea that the outcome of any 
hard scattering process is determined by the 
initial scattering process among elementary constituents -- the
quarks and gluons -- independently from the hadronic phase  
of the reaction, is a well rooted concept in our current view
of high energy phenomena. 
The cross sections for both inclusive and semi-inclusive hadronic processes 
factor out into a ``short distance'' 
($<<$ hadronic size) perturbatively 
calculable part, and a ``large distance'' ($\approx$ hadronic size) measurable 
part that is directly related to the quarks and 
gluons distribution inside the 
hadrons.
During the scattering process the partons are essentially
``free'' -- modulo perturbative-QCD (pQCD) radiation  -- the 
trasmogrification of partons into hadrons and vice-versa
happening at a much too large time scale to influence the outcome
of the reaction.  
Essentially, all hadronic reactions, from $e^+e^- \rightarrow$ hadrons, to 
Deep Inelastic Scattering (DIS), to high energy hadron-hadron reactions
are interpreted using this concept.
Another aspect of duality, Bloom and Gilman (BG) duality \cite{BG}, is the observation
that
by lowering the center of mass energy of the hard scattering process, 
{\it i.e.} by considering the production of resonances, 
the cross section follows in average a curve similar to the
DIS one. The implications of BG duality are twofold: 
on one side the production of resonances seems to be still 
influenced by partonic degrees of freedom; on the other,  
the correlation functions encoding the non-perturbative 
structure of the proton might have a common 
origin with the ones in the resonance region.  

A particularly interesting result was found 
in studies of inclusive reactions with no hadrons in the 
initial state, such as $e^+e^- \rightarrow$ hadrons, and 
hadronic $\tau$ decays \cite{SHI}. It was  
pointed out that, because of the truncation of the PQCD 
asymptotic series, terms including 
quark and gluon condensates play an increasing role as the center 
of mass energy of the process decreases.
Oscillations in the physical observables were then found
to appear if the condensates are calculated in an instanton 
background. Such oscillating structure, calculated in \cite{SHI} 
for values of the center of mass energy above the 
resonance region, is  
damped at high energy, hence warranting the onset of parton-hadron
duality. 
%Interesting information on the structure of the QCD vacuum.
%that is they are extended to values where PQCD is expected to be valid, 
%(down to a scale $Q_o^2 \approx 1 $ GeV$^2$ $<< \Lambda_{QCD}^2$).  
In this contribution we examine a related question,
namely whether it is possible to extend the picture of duality 
explored in the higher $Q^2$ region \cite{SHI}, to the resonance region, or to the BG domain. 
A necessary condition is to determine 
whether the curve from the perturbative regime smoothly 
interpolates through the resonances, or violations of this
correspondence occur.
The latter would indicate that we are entering 
a semi-hard phase of QCD, where preconfinement effects \cite{BCM} are present.

%%%%%%%%%%%%%%%%
\begin{figure}[ht]
\epsfxsize=6cm   %width of figure - will enlarge/reduce the figures
\centerline{\epsfxsize=5cm\epsfbox{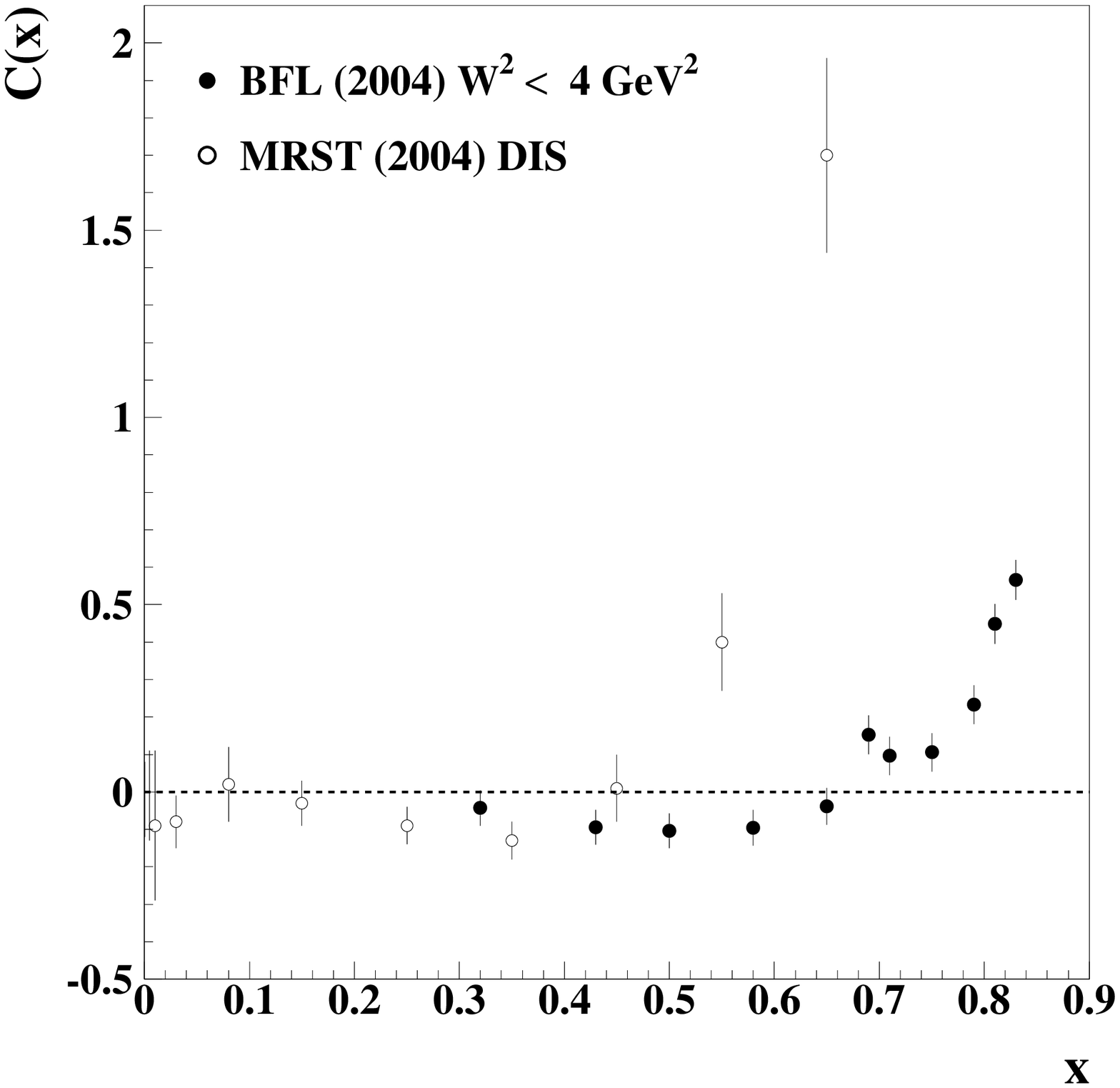}}
\centerline{\epsfxsize=5cm\epsfbox{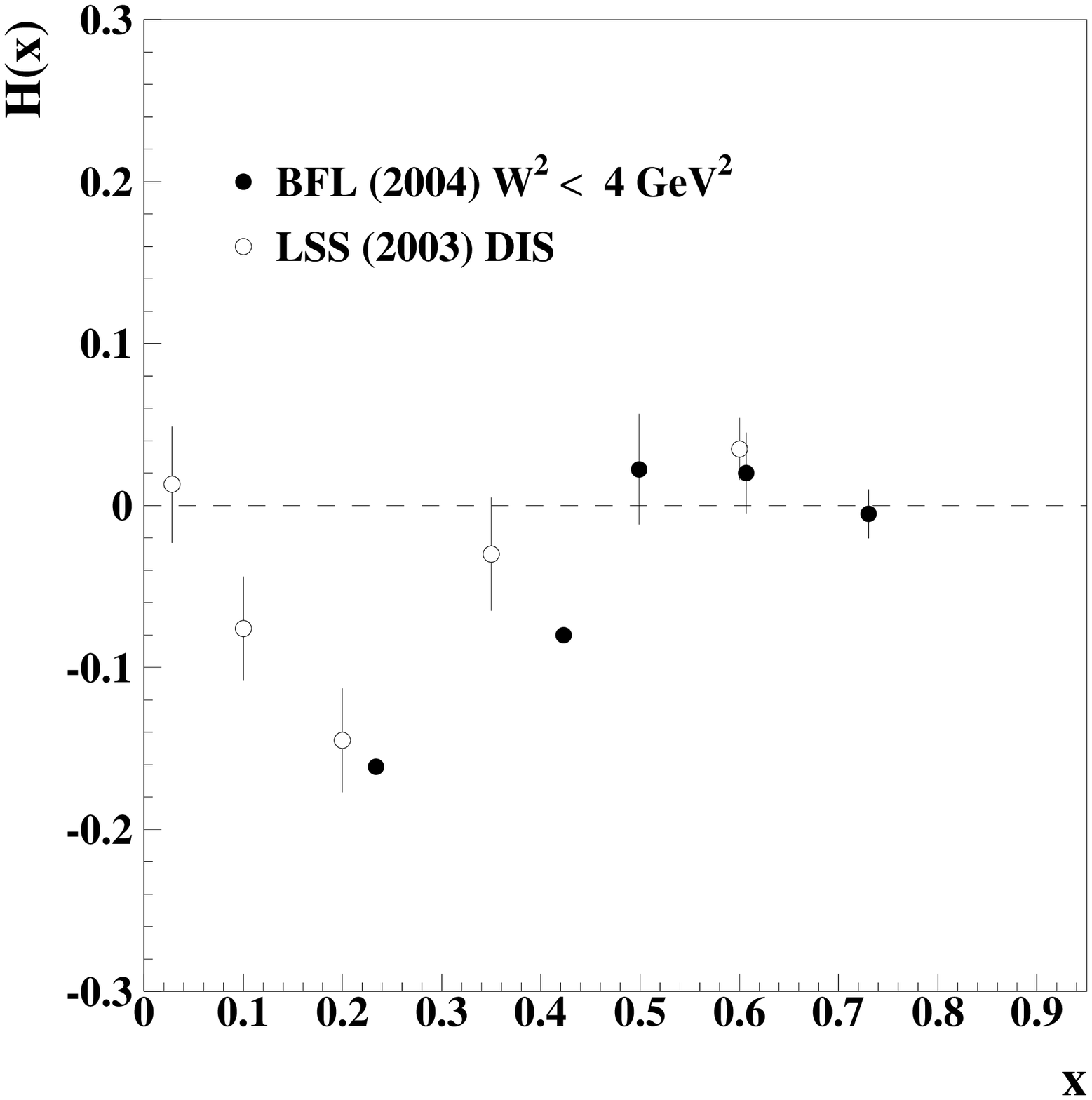}}   
\caption{Comparison of HT contributions for both the 
structure function $F_2$ (upper panel) and the polarized structure
function $g_1$ (lower panel)
in the DIS and resonance regions, respectively. The full circles are 
the values obtained 
in the resonance region \protect\cite{BFL1}.  
For $F_2$ these are compared with extractions using DIS data, 
from different collaborations: 
MRST (unpolarized) \protect\cite{MRST04}. For $g_1$ they are compared to 
the extraction from \protect\cite{Sidorov}. 
Notice that we show our results 
in a factorized model for $F_2$, and in a non-factorized one for $g_1$ for 
a consistent comparison with \protect\cite{VM,MRST04,Sidorov}.}
\label{unp_cht}
\end{figure}

%%%%%%%%%%%%%%%%%%%%%%%%%%%%%%%%%%%%%%%%%%%%%%%%%%%%%%%%%%%%%%%%%%%%%%

\section{Breakdown of Factorization}
BG duality is considered to be fulfilled 
if the extrapolation 
using PQCD evolution from high 
$Q^2$ and $W^2$ into the resonance
region, agrees with the experimental data in this region.
The accuracy of current data   
allows us to address the question of 
{\em what extrapolation from the large $Q^2$, or asymptotic
regime 
the cross sections in the resonance region should be compared to}.
In order to best address possible 
ambiguities in the analyses due 
the choice of an ``averaging procedure'' for 
the data in the resonance region, one should consider 
the following complementary methods:  
\begin{subequations}
\label{average}
\begin{eqnarray}
& & I(Q^2)  =  \int^{x_{\mathrm{max}}}_{x_{\mathrm{min}}} 
 F_2^{\mathrm{res}}(x,Q^2) \; dx 
\label{Ires}
\\ \nonumber \\
& & M_n(Q^2)   =  {\displaystyle \int_0^1} dx \, \xi^{n-1} \, \frac{F_2^{\mathrm{res}}(x,Q^2)}{x} \, 
\, p_n 
\label{moment}
\\ \nonumber \\
& & F_2^{\mathrm{ave}}(x,Q^2(x,W^2))  =  F_2^{\mathrm{Jlab}}(\xi,W^2) 
\label{F2ave}
\end{eqnarray}
\end{subequations}
where $F_2^{\mathrm{res}}$ is evaluated using the experimental data  
in the resonance region
%%%%
\footnote{Similar formulae hold for the polarized structure function, $g_1$.}.
%%%%
In Eq.(\ref{Ires}), for each $Q^2$ value: 
$x_{\mathrm{min}}=Q^2/(Q^2+W_{\mathrm{max}}^2-M^2)$, and 
$x_{\mathrm{max}}=Q^2/(Q^2+W_{\mathrm{min}}^2-M^2)$. 
$W_{\mathrm{min}}$ and $W_{\mathrm{max}}$ delimit either 
the whole resonance region, {\it i.e.}
$W_{\mathrm{min}} \approx 1.1$ GeV$^2$, and $W_{\mathrm{max}}^2 \approx 4$ GeV$^2$, 
or smaller intervals within it.
In Eq.(\ref{moment}), $\displaystyle \xi$ 
is the Nachtmann 
variable \cite{NAC}, and $M_n(Q^2)$ are Nachtmann moments \cite{NAC};
$\displaystyle p_n$ is a kinematical factor \cite{NAC}.
The r.h.s. of Eq.(\ref{F2ave}), $\displaystyle F_2^{\mathrm{Jlab}}(\xi,W^2)$, 
is a smooth fit to the resonant data \cite{jlab_exp}, valid
for $1< W^2 <4$ GeV$^2$; 
$F_2^{\mathrm{ave}}$ symbolizes the
average taken at the $\displaystyle Q^2 \equiv (x,W^2)$ of the data.

Besides ambiguities in the averaging procedure, 
in principle any extrapolation 
from high to low $Q^2$ is fraught with 
theoretical uncertainties ranging from the propagation of 
the uncertainty on $\displaystyle \alpha_S(M_Z^2)$
into the resonance region
to the appearance of different types
of both perturbative and power corrections in the low $Q^2$ regime. 
A program to address quantitatively these sources of theoretical errors
was started in \cite{SIMO1,BFL1}.
In this contribution we present results on the extraction of 
the dynamical Higher Twist (HT) terms from the resonance region,  
and we compare them to results obtained in the DIS region 
\cite{VM,MRST04,Sidorov}. A clear discrepancy marking perhaps a breakdown of
factorization at low values of $W^2$ is seen for the unpolarized structure 
function, $F_2$ (upper panel). More data at large $x$ are needed in order to 
draw conclusion for the polarized structure function, $g_1$. 

An obvious conclusion is that  
in correspondence of the most prominent resonances,
we enter the non-perturbative regime.
%The corresponding space-time picture is that 
%due to the smallness of $W^2$,
%the current fragments into a jet
%with a process duration, $\tau$, that 
%is comparable to the hadrons formation time, $\tau_f$.
The ``snap-shot'' picture of the proton's 
pointlike partonic configurations is replaced 
with a ``blurred'' image  
that encompasses a range of distance scales.  
Yet, the difference between the 
PQCD-based extrapolation from large $Q^2, W^2$ 
and the smooth average of the resonances can be considered
to be small, as quantified by us in Fig.1.
This motivated one of us (S.L.\cite{theory1}) to model 
the low $W^2$ region 
by considering modified evolution equations along the line of the   
Bassetto-Ciafaloni-Marchesini (BCM) \cite{BCM} equations that 
generate ``preconfinement'' 
in the hadronic phase of high energy processes. 
BG duality and its violations, can then be explained in 
terms of the mass distribution of Color Neutral clusters of quarks
and gluons that characterize the semihard phase. 
    
%%%%%%%%%%%%%%%%%%
\section{Conclusions and Outlook}
%For hadronic scattering processes in an intermediate region 
%characterized by low values of $W^2$ there exists a region where 
%neither PQCD, nor chiral perturbation theory, {\it i.e.} 
%the low energy effective theory of strong interactions, seem
%to be able to quantitatively reproduce the experimental data.
We investigated the phenomenon of parton-hadron duality for 
inclusive unpolarized and polarized $ep$ scattering. 
Our conclusions are as follows:
{\it i)} Bloom and Gilman duality can now be studied quantitatively, 
because of the increased accuracy of the data;
{\it ii)} We interpret the ``apparent agreement'' between data and the 
pQCD curve in the resonance region as a signature of a breakdown of the
twist expansion at low $W^2$. The HTs extracted from the resonance region
are in fact both in qualitative and quantitative disagreement with the ones 
extracted from DIS;
{\it iii)} BG duality needs to be treated distinctively from
parton-hadron duality at higher $Q^2$ and center of mass energy values.
The $Q^2$ dependence in the Few GeV region can be modeled by considering 
the preconfinement property of QCD, as a hybrid phase where 
clusters of color connected partons interact directly with the probe.
 
As an outlook, the 12 GeV program at Jefferson Lab, will enable us
to validate the picture behind duality and its different manifestations, 
by addressing quantitatively a variety of reactions: 
from polarized scattering at large $x$ and $W^2$, 
to semi-inclusive experiments ...
%\section*{Acknowledgments}
%This is where one acknowledge funding bodies etc.  

\section*{Acknowledgments}
This work is supported by the U.S. 
Department
of Energy grant no. DE-FG02-01ER41200.

\end{document}